\documentclass[hidelinks, conference]{IEEEtran}
\IEEEoverridecommandlockouts
\usepackage{cite}
\usepackage{amsmath,amssymb,amsfonts}
\usepackage{algorithmic}
\usepackage{graphicx}
\usepackage{textcomp}
\usepackage{xcolor}
\def\BibTeX{{\rm B\kern-.05em{\sc i\kern-.025em b}\kern-.08em
    T\kern-.1667em\lower.7ex\hbox{E}\kern-.125emX}}

\usepackage[OT1]{fontenc}
\PassOptionsToPackage{hyphens}{url}
\usepackage{hyperref}
\usepackage[inline]{enumitem}
\usepackage{cleveref}
\usepackage{pgffor}
\usepackage[acronym]{glossaries}
\usepackage{glossaries-extra}
\usepackage{cleveref}
\setabbreviationstyle[acronym]{long-short}

\begin{document}

\title{An OPC~UA-based industrial Big Data architecture
\ifdefined\commitsha \\\vspace{3mm}\footnotesize 
    Document Version: v\piplineid-\commitsha\fi}
\newacronym{api}{API}{Application Programming Interface}
\newacronym{as}{AS}{Address Space}
\newacronym{cdns}{CDNs}{Content Delivery Networks}
\newacronym{da}{DA}{Digital Assistant}
\newacronym{dr}{DR}{Device Registry}
\newacronym{ec}{EC}{Edge Computing}
\newacronym{eai}{EAI}{Enterprise Application Integration}
\newacronym{erp}{ERP}{Enterprise Resource Planning}
\newacronym{etl}{ETL}{Extract Transform Load}
\newacronym{ie}{IE}{Information Engine}
\newacronym{re}{RE}{Reference Architecture}
\newacronym{iot}{IoT}{Internet of Things}
\newacronym{it}{IT}{Information Technology}
\newacronym{im}{IM}{Information Model}
\newacronym{ods}{ODS}{Operational Data Store}
\newacronym{opcim}{OPC~UA~IM}{\gls{opcua} Information Model}
\newacronym{opcqm}{OPC~UA~QM}{\gls{opcua} Query Model}
\newacronym{ml}{ML}{Machine Learning}
\newacronym{mes}{MES}{Manufacturing Execution System}
\newacronym{mom}{MOM}{Message Oriented Middleware}
\newacronym{opcua}{OPC~UA}{Open Platform Communications Unified Architecture}
\newacronym{odm}{ODM}{Object Document Mapper}
\newacronym{orm}{ORM}{Object Relational Mapper}
\newacronym{ot}{OT}{Operational Technology}
\newacronym{plc}{PLC}{Programmable Logic Controller}
\newacronym{qm}{QM}{Query Model}
\newacronym{rt}{RT}{Retrieval Template}
\newacronym{rdbms}{RDBMS}{Relational Database Management System}
\newacronym{scada}{SCADA}{Supervisory Control and Data Acquisition}
\newacronym{sam}{SAM}{Simple Asset Management}
\newacronym{sme}{SME}{small and medium-sized enterprises}
\newacronym{soa}{SOA}{Service Oriented Architecture}
\newacronym{sparql}{SPARQL}{SPARQL Protocol and RDF Query Language}
\newacronym{tsn}{TSN}{Time Sensitive Networking}
\newacronym{ttl}{TTL}{Time To Live}
\newacronym{uml}{UML}{Unified Modelling Language}
\newacronym{xml}{XML}{Extensible Markup Language}
\newcommand{\current}[0]{
  \vspace{5mm}
  \textcolor{red}{Current Position\hrule}
  \vspace{3mm}
}

\newcommand{\todoEnum}[1]{
  \vspace{5mm}
  \textcolor{red}{
    ToDo:
    \begin{enumerate}
      \foreach \todoitem in {#1} {\item \todoitem{}}
    \end{enumerate}
  }
}

\newcommand{\todo}[1]{
  \textcolor{red}{
    ToDo: \foreach \todoitem [count=\i] in {#1} {\i) \todoitem{} }
  }
}

\newcommand{\tdref}[0] {\textcolor{red}{[$\forall$ref]}}

\newcommand{\sthu}[1]{\marginpar{\tiny\textcolor{violet}{StHu: #1}}}
\newcommand{\edhi}[1]{\marginpar{\tiny\textcolor{blue}  {EdHi: #1}}}
\newcommand{\siho}[1]{\marginpar{\tiny\textcolor{orange}{SiHo: #1}}}
\newcommand{\maur}[1]{\marginpar{\tiny\textcolor{orange}{MaUr: #1}}}

\Crefname{figure}{Fig.}{Figs.}

\author{\IEEEauthorblockN{Eduard Hirsch, Simon Hoher, Stefan Huber}
\IEEEauthorblockA{JR Centre for Intelligent and Secure Industrial Automation,
Salzburg University of Applied Sciences, 5412 Puch, Austria\\
\{eduard.hirsch,simon.hoher,stefan.huber\}@fh-salzburg.ac.at}}

\maketitle

\begin{abstract}
Industry 4.0 factories are complex and data-driven. Data is yielded from many
sources, including sensors, PLCs, and other devices, but also from IT, like ERP
or CRM systems. We ask how to collect and process this data in a way, such that
it includes metadata and can be used for industrial analytics or to derive
intelligent support systems. This paper describes a new, query model based
approach, which uses a big data architecture to capture data from various
sources using OPC~UA as a foundation. It buffers and preprocesses the
information for the purpose of harmonizing and providing a holistic state space
of a factory, as well as mappings to the current state of a production site.
That information can be made available to multiple processing sinks, decoupled
from the data sources, which enables them to work with the information without
interfering with devices of the production, disturbing the network devices they
are working in, or influencing the production process negatively. Metadata and
connected semantic information is kept throughout the process, allowing to feed
algorithms with meaningful data, so that it can be accessed in its entirety to
perform time series analysis, machine learning or similar evaluations as well as
replaying the data from the buffer for repeatable simulations.
\end{abstract}


\section{Introduction}
\label{sec:intro}

\subsection{Motivation}
\label{subsec:motivation}

In Industry 4.0, data is a critical resource that is used to optimize and
control manufacturing processes and reach for continuous improvement based on
industrial analytics. Industry 4.0 is data-driven \cite{klingenberg2021industry}
and in order to fully understand the produced data, it is necessary to fully
manage and understand base and config data \cite{hofmann2016digitale} of the
shop floor, but also from related \gls{it} systems providing, e.g., order or
inventory information. They need to be precisely defined, eliminating gaps
created by missing environmental information or only known or filled in by human
experts. Thus, we ask \textit{first} how to provide a holistic well-defined data
view of the production environment and in which state the production is
currently in? \textit{Second}, in what way should the data be collected from
various sources and read as often necessary, without breaking their integrity,
or disturbing the production process but decreasing load on them as well as on
the network? \textit{Third}, what is an appropriate way harmonize data, as well
as to let accessing sinks understand data semantically?

For common \gls{it} infrastructure the process of collecting data, storing and
analyzing is well established and has been studied and deployed for years.
Various (big) data tools and techniques \cite{warren2015big}, storage concepts
\cite{armbrust2021lakehouse} and data analysis methods, e.g., using data
warehouse techniques \cite{kimball2011data}, are applied. However, data produced
in \gls{ot} and Industry 4.0 setups lack the well established techniques for
fetching and storing it in a way that is comparable to IT environments. Thus, it
would be desirable providing the same level of analytic abilities or to derive
and feed support systems, in other words, to allow multiple and extendable
processing sinks.

\gls{it} and \gls{ot} are related, and concepts of the \gls{it} domain are often
transferred to \gls{ot}, e.g., the concept of \gls{iot} or \gls{ec}. Although,
the basic ideas are similar to \gls{it}, perspectives in \gls{ot} are different
and require adjusted approaches to what are typically used in \gls{it}
environments \cite[table 2-1]{stouffer2015}. So \gls{ec} was developed for IT to
bring data sources closer to end-users, thus, multiple \gls{ec} devices in
proximity to the end-users. For \gls{ot} environments \gls{ec} devices are  
brought on premise, to the data sources on shop floor.
\cite{sitton2020edge}.

Like \gls{ec} techniques differ when applied in \gls{it}, collecting information
from various sources and provide them for analysis defer from \gls{ot} methods.
A solution that comes close to requirements of \gls{ot} environments are the
various \gls{iot} solutions, provided by different vendors like Azure, AWS or
Google Cloud, that are designed to collect data from \gls{iot} devices and
provide it for analysis. However, these solutions are in many cases not suitable
to be used in \gls{ot} environments, which has a couple of reasons. First,
collecting data from a \gls{plc} requires a linkage from the devices to the
cloud which prevents a network from being sealed off. Data is transferred to
a central server, not under the control of the company itself. Even with
security mechanisms \cite{abdulsalam2022security} applied, it is never as secure
\cite{hou2020industrial}\cite{ettredge2018trade} as compartmentalizing the
network, causing issues with data privacy or integrity. Second, reliability is a
concern, because network bandwidth or latency can be a major issue in \gls{ot}
\cite{bali2013effect} causing data to be delayed or even lost. Third, regulatory
requirements and compliance issues may also prevent the use of cloud solutions
in certain \gls{ot} industries. Similar issues apply to ready-to-use edge
devices which come with pre-install software and services. These devices
decrease the need for a central server but are, due to their pre-install
software and services, also not entirely trustworthy.

Further, the \textit{three V's} of big
data~\cite{laney01controlling3v}\cite{warren2015big} are a major challenge for
\gls{ot} environments and need to be considered when designing a solution, even
when production sites are smaller like for \gls{sme}. A solution needs to be
able to handle a large \textit{variety} which refers to the diverse data types
that are generated by different sensors, machines, and systems on the shop
floor, such as (semi-)structured data (e.g., machine readings, sensor data) and
unstructured data (e.g., images, text). The \textit{volume} is, when working
with real-time data, also a major issue. Consider just a single measurement
point with one kilobyte and a sampling rate of $1$Hz ($1$KB/s), kept for
analysis for one year. That leads to a data volume of $84.38$MB per day, around
$2.47$GB per month and $30.08$GB per year, for a single data point. Finally,
\textit{velocity} is another important aspect, because data generated in
real-time, needs to be queried, stored and processed in real-time.

Focusing on the initial data collection aspects and on the similarities between
\gls{it} and \gls{ot} environments, we are aiming for a solution that is
comparable to IT environments based on big data techniques\cite{warren2015big}.
These techniques are introduced in order to handle data at a single point of
truth applying location and access transparency. The well established protocol
and information modelling language \gls{opcua} is recommended by RAMI 4.0
\cite{zvei2015} and a standardized communication protocol. It is used in
industrial automation and control systems to enable secure and reliable exchange
of data between different devices and systems. Therefor, it is a capable basis
for reading information from industrial devices on shop floor. But we note that
we are confronted with challenges that are specific to \gls{ot} environments.
The presented solution is intended to be placed on site, where edge computing is
explicitly out of scope and the final system completely under the control of the
company and not dependent on third party vendors.

\subsection{Contributions}
\label{subsec:contributions}

In this paper we pursue the following methodology: We start by deriving the
requirements for a Big Data architecture that is applicable to OT environments
and, in particular, we identify obstacles that are specific to OT. Then the
state of the art of corresponding IT architectures is reviewed. The most
favorable IT-architecture is picked and then enhanced to address the OT aspects
previously identified. By combining these approaches, a comprehensive
architecture was developed that can act as a reference for a Big Data
architecture in OT environments. It acquires, processes and persists data, which
is a prerequisite for analyzing data. This allows modular downstream
applications like building digital support systems, which are able to reason
based on the available data. This information system we are naming \gls{ie}.
There are four novel contributions we want to highlight:

\textit{First}, this paper offers a \gls{qm} approach to clearly define
which devices are needed to be monitored and what data is needed to be
collected. The goal is to provide a holistic view (state space) of the
production environment by consistently and persistently collecting data and
metadata from various devices and systems without breaking their integrity.

\textit{Second}, the paper shows how to decouple data sources from data sinks,
to reduce the load on the network, the computational pressure on the information
sources, as well as the effort for dealing with historic data on the information
sources. This increases flexibility and allows for reading data, at any time, in
definable granularities and formats, as well as to access historical data at and
from any time. Value-added services downstream this proposed architecture, are
operated besides the OT system and shall not put too much computational or
communication burden on the existing system.

\textit{Third}, a data flow concept is discussed, which makes it possible to
transform, harmonize data from different sources and to augment data with
additional information and metadata. This is done in a generic way, so that
various transformation engines can be used, depending on the data type or
format.

\textit{Fourth}, an architecture, the \gls{ie}, is described, combining these
novel approaches into a system having standardized \gls{opcua} interfaces. It is
offering data in various formats and allowing to use the data in different ways.
The main goal here lies on feeding analytical and \gls{ml} processes, which need
data to be preprocessed or augmented in order to be digested directly. Thus,
there should not be the requirement of heavy additional preprocessing for actual
analysis, training/inference (\gls{ml}), as well as for digital shadows or
twins needing automatic information inflow according to\cite{Kritzinger2018}.

\gls{scada} and \gls{mes} system are central components of \gls{ot}
environments, and need to be considered when designing solutions for data
processing systems, because they are already in proximity to the shop floor.
They collect and work with controlling information or devices and thus, can
provide essential information that are vital for doing industrial analytics or
creating support systems. Regardless, it is important to clarify that these
systems are designed for monitoring and controlling the production
process~\cite{meyer2009manufacturing}. The presented solution is not intended to
replace existing \gls{scada} or \gls{mes} systems, but to complement them.
\gls{scada} and \gls{mes} are regarded as a valuable source of data, used to
feed the presented system.

The solution considered in this paper is designed to keep the data on site
without the need of transferring collected information to a remote server. It is
discussed outside the context of edge computing. We are targeting architectural
requirements for a system, that is completely under the control of the company,
and allows the exchange of software products as well as is not dependent on
specific third party vendors. The on-site solution is considered to be used in
\gls{ot} environments, in proximity to where the data is generated.

\subsection{Prior and related work}
\label{subsec:related}

\begin{figure}[t]
  \centering
  \includegraphics[width=0.48\textwidth]{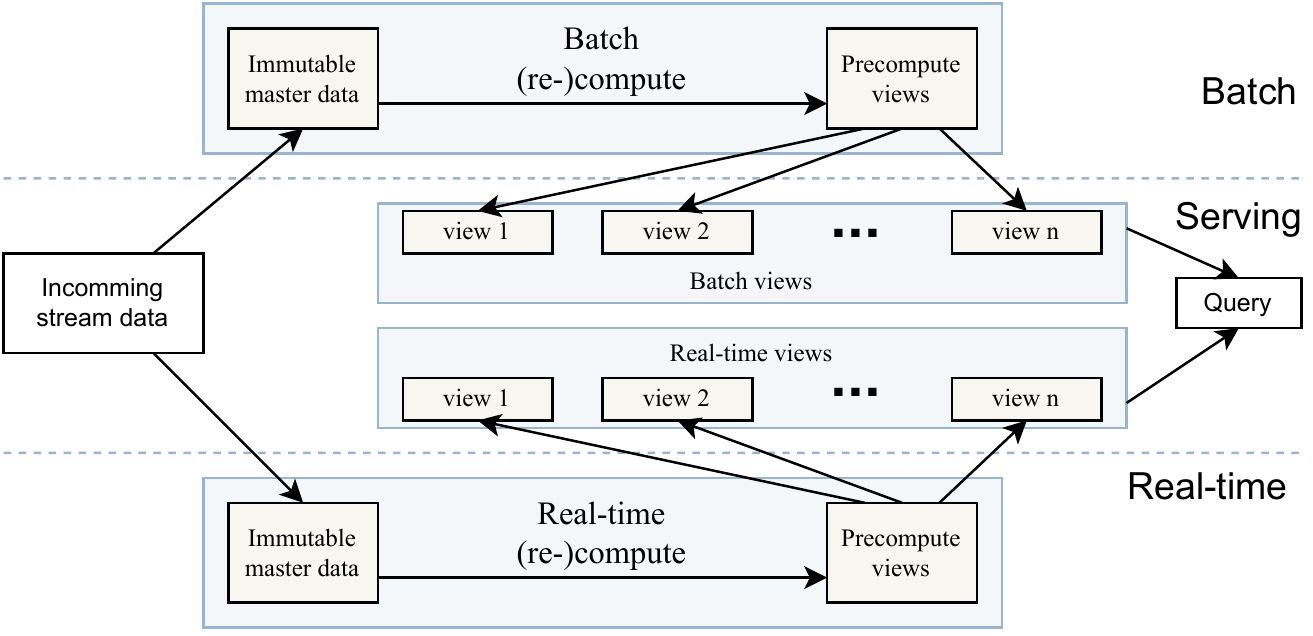}
  \caption{Lambda Architecture~\cite{warren2015big}, a reference architecture
  for implementations of big data systems.}
  \label{fig:lambda-arch}
\end{figure}

\gls{opcua} as a communication protocol widely used in industrial automation and
control systems acts as a bridge between \gls{it} and \gls{ot}. It's providing a
standard way for different devices, systems, and applications to communicate
with each other, embracing a \gls{soa}\cite{openGroup2009}\cite{mahnke2009opc}
approach. \gls{opcua} enables interoperability, security, scalability, and
flexibility, making it suitable for various industries and applications. It
helps organizations to integrate different technologies and systems across their
operations and create efficient and secure systems that leverage the benefits of
both \gls{it} and \gls{ot} domains. That are the reasons, it has been chosen as
a base communication protocol in the \gls{re}. But just reading information from
devices is not enough, because the data needs to be processed and stored in a
way, that it can be used for further analysis or real-time inference. A starting
point for an architecture, which is presented in literature for processing and
analyzing large amounts of data, in distributed computing environments, is the
Lambda and the Kappa Architecture~\cite{warren2015big}. The Lambda Architecture
shown in \Cref{fig:lambda-arch} is a combination of two architectures: a batch
and a stream processing part. Data is processed immediately from a stream or in
large quantities at once respectively. Results of the computations from the both
parts are stored in a separate storage area and can be accessed by users or
other systems. Depending on the requirements of the use case, Lambda
Architecture can be adapted to the Kappa Architecture~\cite{kreps2014kapa}. The
Kappa Architecture simplifies the architecture by removing the batch processing
part and handling data using the paradigm "everything~is~a~stream". Kappa is
compensating for batch processing by facilitating modern streaming architectures
and streaming historical data~\cite{lin2017lambda}. We will use Lambda and Kappa
Architecture for building an information system, that is able to process data
from various sources in industrial environments.

At public cloud providers, like Azure, AWS and Google, we can see \gls{iot}
solutions, applying the Lambda/Kappa Architecture demonstrated in
\cite{cloud2023} and addressing parts of the issues mentioned in
\cref{subsec:motivation}. There are also similar implementations available based
on open source software and on-premise software \cite{cakir2022enabling}.
Nevertheless, they are very general in terms of their usage and application or
do not fully cover the requirements of an industrial environment
\cite{survey2018} like discussed in \cref{subsec:motivation}.

\section{The OPC~UA Query Model}
\label{sec:opcuaquerymodel}

In this section, we will describe a \gls{qm} on which data retrieval of the
presented architecture in \cref{sec:re} is based on. It is a method to retrieve
data from multiple \gls{opcua} servers in a structured way. The \gls{qm}
itself is based on a \gls{opcim} and used to define which data points (states)
to query from \gls{opcua} servers. The \gls{opcim} is a hierarchical structure
and offers information modelling in a standardized way, making it easier for
different devices and systems to exchange information. It defines various node
classes, which can represent real-world or virtual entities with their
properties. These nodes have unique identifiers, are organized into namespaces
and can have attributes such as data type, value, and access level. It is
providing a common language and structure for representing data and information
and can be made available over the \gls{opcua} \gls{as} \cite{mahnke2009opc}.

\begin{figure}[t]
    \centering
    \includegraphics[width=0.45\textwidth]{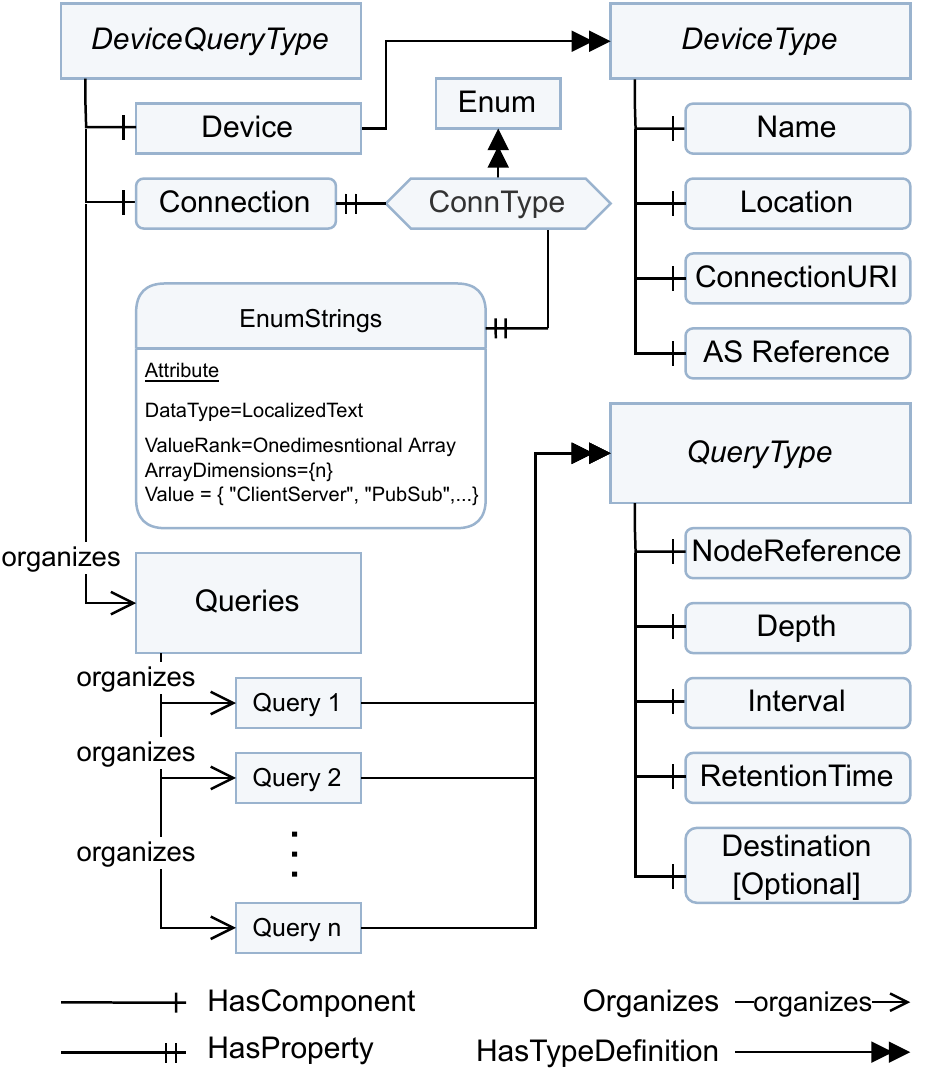}
    \caption{\gls{opcim} for query model based queries}
    \label{fig:querymodel}
\end{figure}

We define an \textbf{\gls{opcqm}} as a \gls{opcim}, that allows to declare for
an \gls{opcua} source server off a device, how data retrieval should be scheduled
and how data points are fetched. \gls{opcqm} defines in what granularities
data is read, together with the information of how long this data is kept.
\Cref{fig:querymodel} illustrates the \gls{opcqm}, based on the Graphical
\gls{opcua} Notation \cite{opcua2019} and is explained in the following.
\textit{DeviceQueryType} defines the type for querying data from a device. An
instance needs to be created for each device, that should be queried. It
contains the \textit{Device}, the \textit{ConnectionType} (Client/Server,
pub/sub, ...) and the actual queries. \textit{Device} is an instance of
\textit{DeviceType} which holds the information about a particular device.

The \textit{DeviceType} keeps track of the \textit{Name}, \textit{Location}
(physical) of the device together with the \textit{ConnectionURI} (\gls{opcua}
URI). Important to note is, that \textit{DeviceType} is holding a reference to
the \gls{opcua} \gls{as} of the device. The reference can be a URI, file
location, or a custom reference where the \gls{as} can be downloaded or fetched.
To understand the querying process we need to look at \textit{Queries} which
organizes instances of the \textit{QueryType}. \textit{QueryType} is the
definition of a query. It contains a \textit{NodeReference} identifying the node
to read data from, in what granularity (\textit{Interval}) and in what
\textit{Depth}. \textit{RetentionTime} and \textit{Destination} are related to
the buffering of the data. \textit{RetentionTime} defines how long the data
should be kept in the buffer. The optional \textit{Destination} defines where
the data should be stored to, which can be, e.g., a topic or a queue name in a
message broker. If the parameter \textit{Destination} is not set, the data will
be stored with the same name as the \textit{NodeReference}. The
\textit{NodeReference} should be a namespace, together with a node id or browse
name to identify the node.

\gls{opcim} is a powerful way of modelling information and data, which was the
reason it has been chosen to model the \gls{opcqm}. It is standardized and
powerful enough, to define the structure of the \gls{qm} and the data retrieval
process. In addition, it is possible to model an \gls{as} which manages and
structures the various \gls{opcqm} instances in order to add further metadata,
or to use it for post-processing.

The \gls{qm} itself has solid advantages over on-demand data retrieval. By using
predefined \gls{qm}s, the complexity of navigating the \gls{opcua} address space
is reduced, making it easier to retrieve information from a huge amount
of devices. Additionally, the use of \gls{qm} can also improve the efficiency
of the information extraction process, as the same \gls{qm} can be used for
multiple identical/similar devices or to bulk query data points from one
device, instead of querying each data point individually. Further, the \gls{qm}
exactly defines what is retrieved and where it is located, which makes working
with the data more clear. Knowing the structure, semantics and amount of the
data beforehand is important to estimate the complexity of the production
process in that facility.

\section{Decoupling sources from sinks}
\label{sec:buffer}

Next we discuss the decoupling of sources from sinks. This is done by
introducing a buffer between the sources and the sinks. The buffer is
responsible for collecting data from the sources and distributing it to the
sinks. It is also responsible for storing data for a certain amount of
time, so that the sinks can fetch the data at their own pace.

\begin{figure}[t]
  \centering
  \includegraphics[width=0.36\textwidth]{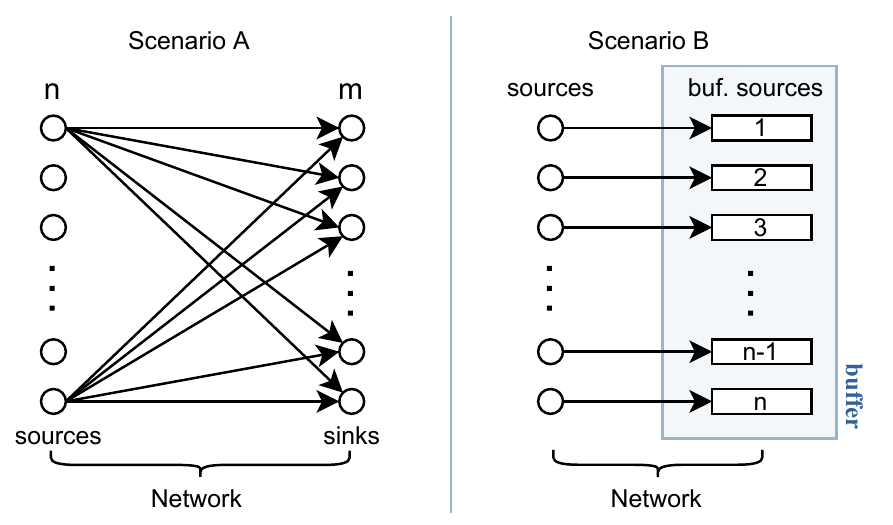}
  \caption{Communication complexity illustration for different scenarios.
  Scenario A: Reading information directly from multiple sinks. Scenario B:
  Replicating sources to a buffer before providing to sinks.}
  \label{fig:complexity}
\end{figure}

Before going into details of the buffer, we first discuss the expected
communication complexity of the system. For transmitting information we assume a
data point of fixed size. When sending that data point, from a source to a sink,
the computational and communication effort is constant, which we express with
the $\mathcal{O}$ notation as $\mathcal{O}(1)$. When reading the same
information from one source, the effort at that source increases by the number
of sinks, thus, assuming $m$ sinks reading the same information causing
$\mathcal{O}(m)$ complexity, as illustrated in \Cref{fig:complexity}, Scenario
A. Sinks keep there computational and communication effort at $\mathcal{O}(1)$.
Concluding further, when declaring $n$ sources and $m$ sinks, the complexity
raises to ${\mathcal{O}(n\cdot m)}$, but on the network only.

In order to reduce information transport complexity, messages and events are
stored in a buffer offering the following advantages. \textit{First}, buffering
of the messages helps to move load from the producing source devices, because as
an intermediate storage it is able to pick up information once and distribute it
to multiple sinks. This intermediate storage leads to $\mathcal{O}(1)$
complexity in terms of communication and local computational effort for reading
information at the \gls{plc} or \gls{opcua} compatible device, respectively.
This is because the information only needs to be fetched once and the rest of
the load is handled by the buffer, like shown in \Cref{fig:complexity}, Scenario
B. Even more, historic data requests do not need to be answered by the sources
anymore at all, but answered instead by the buffer directly. \textit{Second},
the buffer may be separated from the network of the data-holding devices, which
reduces the load on the network of the shop floor. \textit{Third}, the buffer is
inherently decoupling data sources from sinks. This makes communication much
easier, because the sinks only need to connect to a single source of
information. Additional, sinks can be added at any time without affecting the
sources or other sinks. \textit{Fourth}, the buffer is able to store data for a
certain period of time, so that sinks can fetch the data at their own pace. This
enables agnostic sinks, allowing them to fetch data at any point in time from
that selected time period. This gives consumers the possibility to fetch data
multiple times in order to run calculations or attach physical/virtual sinks,
starting simulations as often as needed.

\section{Harmonizing and publishing data points}
\label{sec:harmonize}

\subsection{Data flow}
\label{subsec:dataflow}

\begin{figure}[t]
  \centering
  \includegraphics[width=0.48\textwidth]{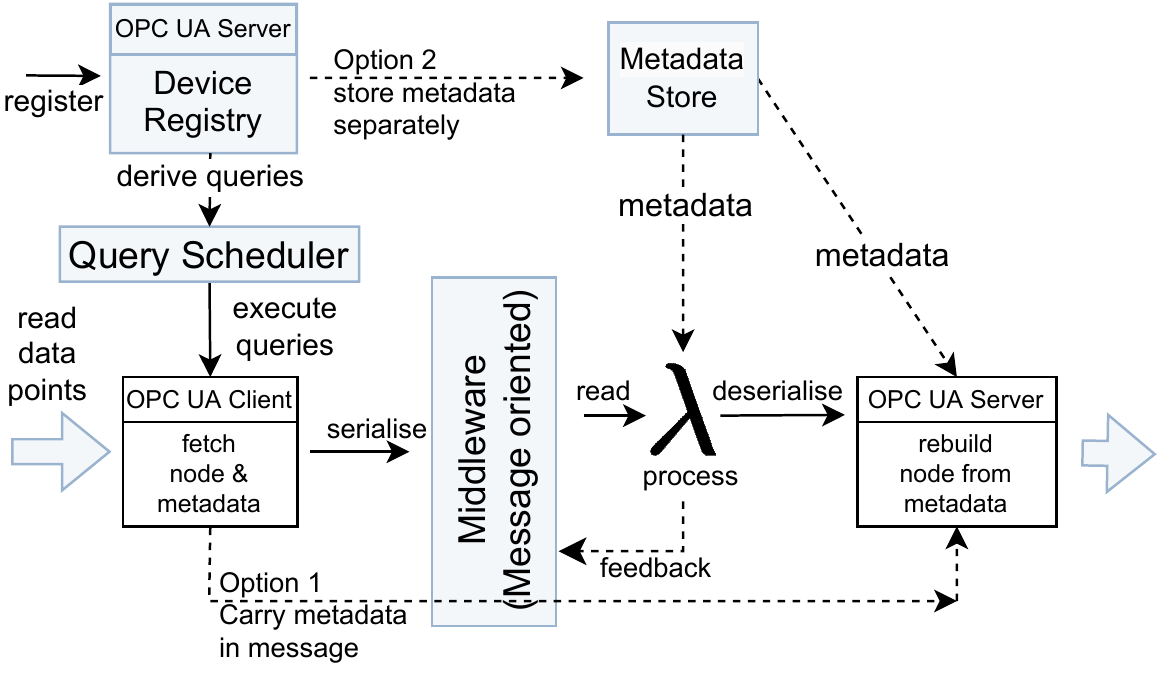}
  \caption{Data flow starting from the fetched data points of a device, till the
  rebuild and publishing for the access of multiple sinks.}
  \label{fig:dataflow}
\end{figure}

In this section the data flow of published data points is described in greater
detail, illustrated in \Cref{fig:dataflow}. First, in order to know what devices
and data points of those devices are relevant, it is necessary to register the
device at a registration authority (\cref{sec:re}). This component is a database
which consolidates information about the devices and their data points as well
as the data model and semantic information using the \gls{opcqm} described in
\cref{sec:opcuaquerymodel}. Once the registration process is done, the system is
able to read the information from the device and store it in a buffer. The
buffer is represented in \Cref{fig:dataflow} as a \gls{mom}. The \gls{mom} has
been chosen here to enable the system collecting various data types and formats,
so that the serialization algorithm can decide how to write the message.

Once the registration with the \gls{opcqm} is done, the system knows which
devices with what data points to fetch in which interval. A scheduler, part of
the connection engine (see \cref{sec:re}), is responsible for reading
from the configured \gls{opcua} node(s). In \Cref{fig:dataflow} there are
two choices emphasized, for handling metadata. The first option is to carry the
metadata with the message itself. The second option is to store the metadata
in a separate database. The first option is more convenient, because the
metadata is directly available. The second option requires an additional query
to a separate storage to get the metadata. However, the second option is more
efficient, because the metadata is only stored once and not repeated for each
message. The second option is also more flexible, because the metadata can be
changed without changing the message itself.

Then the node data needs to be serialized to a format which can be stored in the
\gls{mom}, which is dealing as a buffer introduced in \cref{sec:buffer}.
The serialization done by the writing process can be done in different
ways. Common options are JSON, XML, CSV, Apache Avro, Protobuf or Apache Parquet
when considering bigger file content.

It is preferable to use serialization formats that are commonly know in \gls{it}
and big data, for being able to handle messages out of the box by typical tools
and frameworks available. This makes it easier to preprocess data as represented
in \Cref{fig:dataflow} with symbol $\lambda$. The transformed data can be either
written back to the \gls{mom} or passed on to the serving stage. Passing on
results in deserializing the message, using the metadata to rebuild the
\gls{opcua} node and publishing it on the \gls{opcua} server, to be requested by
clients. If necessary, $\lambda$ is able to read or change metadata stored in
the metadata store (Option 2) or from the message itself (Option 1).

Finally, it is important to note that the buffer is not only able to collect
data from \gls{opcua} sources, but also from other sources. To achieve this, it
is necessary to add another component which is converting other sources to
\gls{opcua} compatible communication partners. Further things to consider for a
buffer are described in \cref{subsec:storage}.

\subsection{Data processing and transformation}
\label{sec:dataprocessing}

Data retrieved based on \gls{opcqm} from the various sources devices and sensors
is provided over \gls{opcua}, thus, well described and structured. However, the
data is often still not suitable for further analysis. For example, the data
might be too detailed or too coarse grained. It might be necessary to aggregate
data points, transform the data into a different format or interpolate missing
time series data. The processing engine $\lambda$ (\Cref{fig:dataflow}) is able
to achieve this in order to make it more suitable for further analysis. It
fetches data from the buffer, processes it and writes the results either back to
the buffer or to a resource for further handling.

In order to understand the state space created by the \gls{opcua} devices
better, it is discussed in the following more detailed. The \gls{opcua} source
devices/servers (e.g. a \gls{plc}, SCADA or ERP systems) providing information,
can be consolidated into a set $O = \{o_1, \dots, o_n\}$.  Each source $o_i \in
O$ provides an \gls{opcua} address space for a particular device. The
\gls{opcqm} defines a subset of nodes from one address space of each device
$o_i$, creating a traced state space $S_i$. The proposed architecture is
responsible for fetching the current state in $S_i$ time-discrete, with the
result of a state trajectory $x_i \in S_i$. A sample $x_i(t)$ is a vector of
data points of the multidimensional $x_i$ indexed with time-discrete timestamps
$t_j$. Concluding, the holistic state space $S$ of the whole environment, can be
declared as the Cartesian product of the traced state spaces $S_i$ of all
sources $o_i \in O$, thus, $S = \prod \limits_{i=1}^n S_i$. The state trajectory
for the whole environment is declared by
$X = \left(\begin{smallmatrix} x_1 \\
\vphantom{\int\limits^x}\smash{\vdots} \\
x_n \end{smallmatrix}\right)$ in $S$.\vspace{0.05cm}

This way of approaching the data produced in the \gls{ot} environment helps to
understand the information space and the data flow. For example, operations on
the holistic view of the environment applied, like identifying correlations
between different data points, creating a base for anomaly detection or
predictive maintenance, are acting in the \gls{re} on a clear defined state
space. This space can be extended or reduced depending on the needs of an
application with a well-defined repeatable output. Furthermore, a solid base is
necessary for harmonizing different timescales \cite{Rangapuram2023} which means
extending, interpolating or aggregating the data points to a commonly shared
timescale. Additionally, normalization in general using various transformations
may utilize cross-device information sharing of different $x_i$.
\section{A Reference Architecture for the Information~Engine}
\label{sec:re}

\subsection{Introduction}
\label{subsec:ieintro}

\begin{figure}[t]
  \centering
  \includegraphics[width=0.48\textwidth]{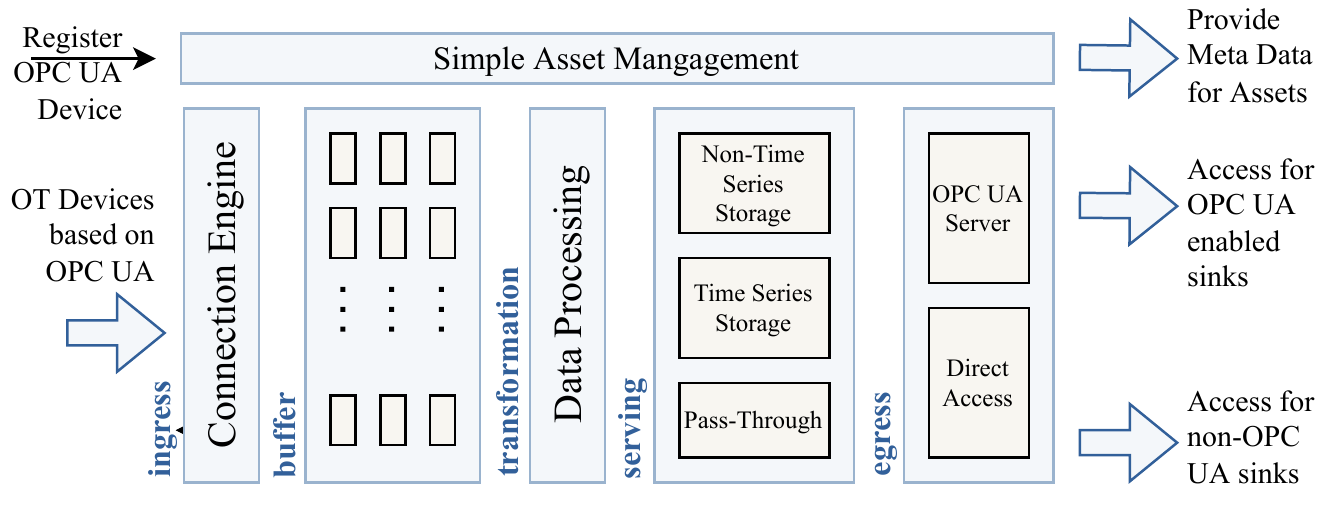}
  \caption{Information processing pipeline based on big data concepts}
  \label{fig:re}
\end{figure}

At this point we propose an information system which puts together the first
three contributions of \cref{subsec:contributions} into a single architecture,
shown in \Cref{fig:re}. We provide a \gls{re} with the main purpose of
collecting data from various sources, buffer it, possibly transform and finally
serve it to other systems, by providing standardized interfaces, together with a
well-defined data model and semantic information. For doing so it is necessary,
that a data source is registered first with the \gls{sam} in order to be able to
collect data from it. The important aspect here is, that the data source is
registered with a \gls{qm}, which defines what data is needed to be collected
together with its granularity. Once the data sources are registered, the
\gls{re} is able to collect data from them and running it through a pipeline,
which is evolved from the Kappa architecture. When looking at \Cref{fig:re} we
can see five main stages, which are described in the following.

\subsection{Pipeline stages}
\label{subsec:pipeline}

\subsubsection{Ingress}
Acts as the entry point. The connection engine components schedule connections
to the \gls{opcua} sources based on the \gls{opcqm} stored in the \gls{sam}. It
fetches the data and publishes it to the buffer. This key component of the
system knows, together with the \gls{sam}, by reading the \gls{opcqm}, which
data is required to be collected. That approach implicitly defines a state space
of the system, thus, conditions the system lives in. These conditions are framed
by the \gls{qm}, helping to understand the size of the system and possibly
derive the complexity of it. Once ingested, the data is handed over to the
buffer.

\subsubsection{Buffer} It deals as an intermediate storage which records the
information on bases of a stream engine. It enables the \gls{re} to read and
write current as well as historic information in real-time. The principles have
been explained in \cref{sec:buffer}.

\subsubsection{Transformation} This is the stage where the data is processed and
modified if necessary. Transformations can be directly applied to the data in
the buffer, but also to the data that has been written already in the serving
stage. Apache Spark, Apache Flink or Kafka Streams are typical examples for
stream processing engines optimized for real-time environments.

\subsubsection{Serving} That stage provides data in an appropriate format for
batch or stream processing sinks. When data is presented to the serving stage it
is already in a format that is suitable for further external processing, and
has been written into it by the processing engine components. The serving stage
is responsible for providing access to the data in a way which is suitable for
the respective application. In case of building a support system this could mean
providing batch access for training \gls{ml}-models, or inferencing in active
systems, applying an already trained \gls{ml}-model using streamed information
retrieved in real-time.

\subsubsection{Egress} The final stage allows access for \gls{opcua} devices or
provides direct access to serving area. Egress stage has been added to harmonize
output of the \gls{re}. It has the aim to provide a standardized way of
accessing information and make the heterogeneity of the data sources transparent
to the outside world. Information can be directly accessed by over the buffer
(pass-through), using a proprietary \gls{api}, or by using the \gls{opcua}
interface. The \gls{opcua} interface provides a standardized way of accessing
the information with the benefit of having a well-defined data model and
semantic information. When using pass-through access it is necessary to query
the \gls{sam} for the data model and semantic information in case needed.

\subsection{Data Storage Add-ons}
\label{subsec:storage}

Storage is a crucial part of the \gls{re} especially for the serving stage. A
lot of information collected from industrial device is stored in regular
intervals which suggest to use time series databases. However, time series
databases are very specific in terms of how to store and access data, which
makes it difficult to store other data types like from \gls{erp} systems.
Classical \gls{rdbms} may also be considered for storing data intermediately,
but they are not designed for scaling well and may have difficulties to deal
with historic access or heterogeneity of the data, because they are quite rigid
in terms of data schema.

At the \textit{serving stage}, data is stored to be used for analysis and
visualization directly without the need of big post-processing steps. Therefor,
data should already fit the requirements for a particular analysis or training
of a \gls{ml} model. The same applies for a \gls{da} which in contrast does
not need a separate storage engine but may directly use the streaming
information from the \gls{re} by facilitating the \textit{pass-through} access.

To describe further database types in the serving stage are highly dependent on
the use case. For example, if the data is used for training a specific
\gls{ml} model, it might be necessary to store the data in a time series
database. These storage engines do not just include traditional databases but
also file storage systems like HDFS or S3, which are ideally distributed and
fault-tolerant and scale well.

Especially when using many techniques performing \textit{batch access}, which
\gls{opcua} is not particularly well suited for, it is beneficial to use files
storage system. This extension, using batch processing, brings the architecture
of the \gls{re} closer to a Lambda Architecture\cite{warren2015big}, which splits
the processing path into a batch and a stream pipeline as explained in
\cref{subsec:related}.

Storing data from the \textit{registering process} can be done in a \gls{rdbms}
or also in a NoSQL database to distinguish actual data from metadata applying
\textit{separation of concerns}. They do not have to meet the requirements of
the buffer, are well known and easier to handle using common frameworks like
\gls{orm} or \gls{odm}.
\section{Conclusion and future work}
\label{sec:conclusion}

The \gls{qm} approach for \gls{opcua} information retrieval has several
advantages over on-demand data retrieval as mentioned in
\cref{sec:opcuaquerymodel}. Together with the architecture of the
\gls{re}, the system design allows for collecting data over a long period of
time, which is necessary for data analysis and gives the possibility to
extrapolate a state space which is a first step towards consistent data with
semantic information for building e.g. value-added services from it, which would
be otherwise difficult to achieve.

The solution presented can be implemented with various technology stacks except
for \gls{opcua} and is compatible with different storage systems, tools and
frameworks in this domain. However, there are still challenges to be solved and
no matter which technology stack is used to implement the \gls{re}, the
different stages of the pipeline need to work seamlessly together, which can be
difficult to achieve. The \gls{re} is a complex system, and it is not
straightforward to integrate the different components as they are often use case
specific, especially when it comes to serving the data.

For the missing support of non-\gls{opcua} data, the \gls{re} can be extended
with intermediate \gls{opcua} servers, used to bridge the gap. This technique is
common in the industry, and fetches information from non-\gls{opcua} sources
which then are made available via an \gls{opcua} server. To make this
integration meaningful, it is necessary to define an \gls{im} and an \gls{as}
for the data to be retrieved, which is not always a straightforward task,
especially filling it with appropriate semantic information.

Future tasks are to implement the \gls{re} in a real-world scenario, to evaluate
the performance for different technology stacks, to improve the system
design, as well as to integrate with industrial analytics.
Further, the \gls{re} is planned to be extended by an additional semantic
layer, which translates \gls{opcua}-based information into semantic languages,
e.g. RDF or OWL based on \cite{ontologyMapping2019}. This will allow for better
semantic search and queries with e.g. \gls{sparql}. Last to mention are efforts
to integrate the \gls{re} with various \gls{ml} systems. These include \gls{ml}
for data streams, federated learning~\cite{li2020a} which includes working with
different geolocations \& \gls{ec} devices as well as approaches for
imitation~\cite{hua2021learning} or inverse~\cite{arora2021survey} reinforcement
learning starting from \cite{schafer2023architecture}.

\bibliographystyle{IEEEtran}
\bibliography{misc/INDIN2023-opcbasedindustrialbigdata}

\end{document}